\documentclass[pra,twocolumn]{revtex4}

\usepackage{amsmath}                          
\usepackage{amsfonts}                         
\usepackage{amssymb}                          

\usepackage{dsfont}                           
\usepackage[usenames]{color}                  

\usepackage{floatflt}                         
\usepackage{graphicx}                         
\usepackage{epstopdf}                         
\usepackage{epsfig}                          

\newcommand{\BibFile}{1d_universal_qwalks}
\newcommand{\BstFile}{1d_universal_qwalks}

\input{Qcircuit.tex}
\newcommand{\ssc}[2][]{^{\vphantom\dagger{#1}}_{#2}}
\newcommand{\calF}{{\mathcal{F}}}
\newcommand{\calG}{{\mathcal{G}}}
\newcommand{\calH}{{\mathcal{H}}}
\newcommand{\etal}{\textit{et~al.}}
\newcommand{\eg}{\textit{e.g.}}
\newcommand\SWAP{\ensuremath{\textit{SWAP\/}}}
\newcommand{\bigO}{{\cal O}}
\newcommand{\<}{\langle}
\renewcommand{\>}{\rangle}

\newcommand{\ajlmultigate}[3]{*+<1em,#3>{\hphantom{#2}} \qw
\POS[0,0].[#1,0];p !C *{#2},p \save+LU;+RU **\dir{-}\restore\save+RU;+RD
**\dir{-}\restore\save+RD;+LD **\dir{-}\restore\save+LD;+LU
**\dir{-}\restore}

\newcommand{\ajlghost}[2]{*+<1em,#2>{\hphantom{#1}} \qw}

\newcommand{\mysymbol}[1]{{\mbox{\raisebox{-0.3em}{\epsfysize=1.2em\epsfbox{#1}}}}}

\newcommand{\zero}{\mysymbol{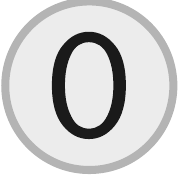}}
\newcommand{\one}{\mysymbol{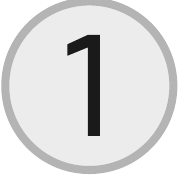}}
\newcommand{\zeroone}{\mysymbol{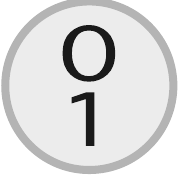}}
\newcommand{\nothing}{\mysymbol{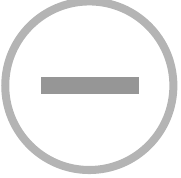}}
\newcommand{\gatecursor}{\mysymbol{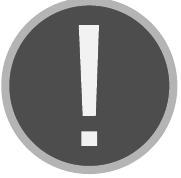}}
\newcommand{\cycle}{\mysymbol{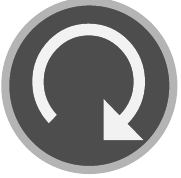}}
\newcommand{\holdcycle}{\mysymbol{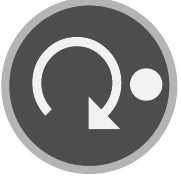}}
\newcommand{\firstboundary}{\mysymbol{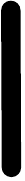}}
\newcommand{\secondboundary}{\mysymbol{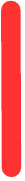}}
\newcommand{\thirdboundary}{\mysymbol{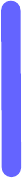}}
\newcommand{\arbitraryqubit}{\mysymbol{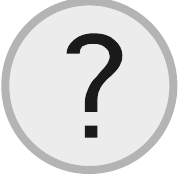}}
\newcommand{\arbitraryqudit}{\mysymbol{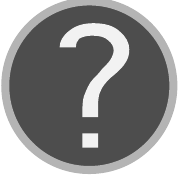}}
\newcommand{\dataA}{\mysymbol{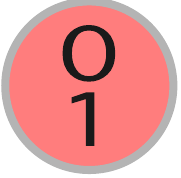}}
\newcommand{\dataB}{\mysymbol{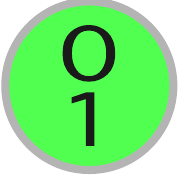}}
\newcommand{\dataC}{\mysymbol{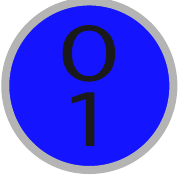}}

\begin{document}


%

\title{Universal quantum walks and adiabatic algorithms by 1D Hamiltonians}

\author{Bradley A. \surname{Chase}}
\email[]{bchase@unm.edu}
\author{Andrew J. \surname{Landahl}}
\email[]{alandahl@unm.edu}
\affiliation{Center for Advanced Studies,
             Dept.\ of Physics and Astronomy,
             University of New Mexico,
             Albuquerque, NM, 87131, USA}



\begin{abstract}

We construct a family of time-independent nearest-neighbor Hamiltonians
coupling eight-state systems on a 1D ring that enables universal quantum
computation.  Hamiltonians in this family can achieve universality either by
driving a continuous-time quantum walk or by terminating an adiabatic
algorithm.  In either case, the universality property can be understood as
arising from an efficient simulation of a programmable quantum circuit.
Using gadget perturbation theory, one can demonstrate the same kind of
universality for related Hamiltonian families acting on qubits in 2D.  Our
results demonstrate that simulating 1D chains of spin-7/2 particles is
BQP-hard, and indeed BQP-complete because the outputs of decision problems
can be encoded in the outputs of such simulations.

\end{abstract}

%
\maketitle


%

\section{Introduction}

With quantum circuits \cite{Deutsch:1989a, Bernstein:1993a, Yao:1993a}, one
can decompose even the most complex quantum computation into a sequence of
simple operations called \emph{gates} that act on simple parcels of
information called \emph{qubits}.  Mathematically, the action of a $T$-gate
quantum circuit on an $n$-qubit pure state can be expressed as $U_T \cdots
U_1|\psi\>$, where $|\psi\>$ represents the input state and each $U_i$
represents a unitary transformation drawn from a quantum gate basis, such as
the universal gate bases described in \cite{Deutsch:1989a, Barenco:1995b,
DiVincenzo:1995a, Lloyd:1995a, Shor:1996a, Kitaev:1997a, Aharonov:1999a,
Boykin:1999a, Boykin:2000a, Shi:2003a, Solovay:2000a, Kitaev:1997b}.  This
mathematical description is usually interpreted as a set of instructions for
applying the gates $U_1, U_2, \ldots, U_T$ in sequence to the qubits.
Feynman noted in Ref.\ \cite{Feynman:1986a} that this description can
instead be interpreted as a \emph{blueprint} for a device in which the gate
sequence is ``printed'' directly into hardware.  Specifically, from the
mathematical description of a quantum circuit, Feynman constructed the
following time-independent Hamiltonian that acts collectively on both the
input state $|\psi\>$ and an auxiliary system he called a ``cursor''
consisting of $(T+1)$ qubits:
\begin{align}
\label{eq:H_feynman}
H_F &= \frac{1}{\sqrt{T+1}} \sum_{t=1}^T U\ssc{t} \sigma_t^+\sigma_{t-1}^- +
     U_{t}^\dagger \sigma_{t-1}^+\sigma_t^-,
\end{align}
where $\sigma^+_k$ and $\sigma^-_k$ denote raising and lowering operators on
the $k$th cursor qubit.  Feynman proved that evolution for a time $T/2$
under $H_F$ with the cursor initialized in the state $|1,0,\ldots, 0\>$ will
efficiently approximate a sequenced implementation of the circuit.  In the
language of quantum information, one would say that Feynman proved that
\emph{continuous-time quantum walks} \cite{Farhi:1997a} are universal for
quantum computation.

Feynman's construction has received little attention as a possible quantum
computing architecture.  This is most likely because a direct implementation
of $H_F$ appears to be far beyond the reach of any known quantum technology.
Some of the difficulties are that $H_F$ has $a)$ four-body interactions
(when two-qubit gates are used), $b)$ spatially nonlocal interactions
(because the cursor gets far away from the data as $T$ and $n$ grow), and is
$c)$ exponentially sensitive to decoherence via Anderson localization
\cite{Landauer:1996a} (because the quantum walk generated by $H_F$ is
effectively on a uniform line).   The central goal of this paper is to
construct a Hamiltonian that achieves the same task as Feynman's but which
is simpler and hopefully closer to a technologically-reachable
implementation.

There is a wealth of previous work on simplifying Feynman's Hamiltonian in
the context of exploring the computational universality of adiabatic
algorithms and of demonstrating the QMA-completeness of finding Hamiltonian
ground states \cite{Kitaev:2002a, Kempe:2004a, Aharonov:2004a,
Oliveira:2005a, Vollbrecht:2007a, Aharonov:2007a}.  (Actually, these efforts
are directed at simplifying a variant of Feynman's Hamiltonian proposed by
Kitaev in \cite{Kitaev:2002a}.) However, these simplifications are done with
the goal of finding Hamiltonians whose \emph{spectra} are close to that of
$H_F$ but which are not necessarily close to $H_F$ \emph{dynamically}.  This
latter kind of closeness is achieved when the \emph{operator norm} of the
difference between the Hamiltonians is close, because, as the following
well-known inequality for Hermetian operators indicates
\cite{Horn:1985a}, this implies that the corresponding dynamics generated
by the Hamiltonians is close:
\begin{align}
\|e^{-iH_Ft} - e^{-iH'_Ft}\| \leq \|H_F - H_F'\|t.
\end{align}

Examples of Hamiltonians whose spectra can efficiently be made close to that
of $H_F$ include a nearest-neighbor Hamiltonian on a two-dimensional grid of
qubits \cite{Oliveira:2005a}, a translationally-invariant nearest-neighbor
Hamiltonian on a line of 30-level systems \cite{Vollbrecht:2007a}, and
nearest-neighbor Hamiltonians on a line of 9-level systems
\cite{Aharonov:2007a}.

Although consideration of Feynman's quantum computing architecture has been
hampered by technological unreachability, it has not been overlooked
entirely.  For example, Margolus \cite{Margolus:1990a} presented an early
generalization in which $H_F$ is replaced by a spatially homogeneous
finite-range 8-body Hamiltonian on a two-dimensional torus of qubits.
Spatial homogeneity makes this model particularly simple---it might best be
termed a \emph{continuous-time quantum cellular automaton}
\cite{Meyer:1996a, Dam:1996a, Watrous:1995a, Shepherd:2006a,
Vollbrecht:2007a}.  However, the 8-body interactions and torus geometry keep
it from being any more practical than Feynman's original architecture.  More
significantly, Margolus' model is only universal for classical, not quantum
computation.  Janzing and Wocjan \cite{Janzing:2005a} improved Margolus'
model so that it is programmable, universal for quantum computation, and
homogeneous on the surface of a cylinder, but at the cost of requiring
10-body interactions among the qubits over a finite range.  Janzing
\cite{Janzing:2007a} has further modified this construction so that with
mild spatial inhomogeneities in the interactions, it achieves programmable
universality using only nearest-neighbor interactions among 3-level systems
on a two-dimensional lattice.  Recently, Nagaj and Wocjan removed these
inhomogeneities in work concurrent to ours \cite{Nagaj:2008a}, and
demonstrated that programmable universality can be achieved by a
continuous-time cellular automaton in one dimension; their construction
requires increasing the number of levels per system from three to ten to
achieve this.

In this paper, we present a programmable, universal architecture for quantum
computation that uses a time-independent nearest-neighbor Hamiltonian on a
one-dimensional ring of 8-level systems.  When combined with the results of
Oliveira and Terhal about the operator norm closeness of Hamiltonians
generated by a certain ``gadget'' perturbation theory technique
\cite{Oliveira:2005a}, we show that we can modify our architecture to use
only qubits at the expense of requiring a two-dimensional rather than a
one-dimensional geometry.  Additionally, we show how our model can also be
used to reduce the state space per carrier from $9$ to $8$ in a recent proof
of the universality of the adiabatic algorithm \cite{Aharonov:2007a}. 

The remainder of our paper is organized as follows.  In Sec.\
\ref{sec:architecture}, we motivate our architecture by defining a relevant
programmably universal quantum circuit family $\calF$.  In Sec.\
\ref{sec:quantum-walk-universality}, we define a family of one-dimensional
8-state Hamiltonians on a ring whose interactions mimic the gates used by
circuits in $\calF$.  In Sec.\ \ref{sec:quantum-walk-universality}, we show
how these Hamiltonians can be used to drive continuous-time quantum walks
that simulate circuits in $\calF$ efficiently.  In Sec.\
\ref{sec:adiabatic}, we demonstrate the universality of our architecture
under adiabatic evolution, following similar arguments to those of Aharonov
\etal\ \cite{Aharonov:2004a, Aharonov:2007a}.  In Sec.\ \ref{sec:qubits}, we
present a method for realizing this model using nearest-neighbor Hamiltonian
interactions between qubits on a regular rectilinear lattice on the surface
of a cylinder by leveraging results of Oliveira and Terhal
\cite{Oliveira:2005a}.  Sec.\ \ref{sec:conclusion} concludes, followed by
Appendix \ref{appendix:example} which steps through how an example 3-gate
simulation would be effected in our model.

%

\section{A programmably universal quantum circuit family}

\label{sec:V-circuit}

%

\subsection{Notation and basic definitions}

We begin by reviewing relevant notation and basic definitions, much of which
can be found in standard textbooks on quantum computation, such as
\cite{Nielsen:2000a,Kitaev:2002a}.  A \emph{qubit} is a quantum two-state
system, the \emph{computational basis} of which we define as $|0\> :=
\binom{1}{0}$, $|1\> := \binom{0}{1}$.  A \emph{gate} is a unitary
transformation on qubits and a \emph{circuit} is a composition of gates.
Gates discussed in this paper include the \textsc{not} \emph{gate}, the
\emph{Hadamard gate}, the \emph{phase gate}, and the \emph{swap gate},
defined in the computational basis as
\begin{align}
X &:=
 \begin{pmatrix}
   0 & 1\\
   1 & 0
 \end{pmatrix},
&
H &:= \frac{1}{\sqrt{2}}
 \begin{pmatrix}
   1 & 1\\
   1 & -1
 \end{pmatrix},
\\[3ex]
S &:=
 \begin{pmatrix}
   1 & 0\\
   0 & i
 \end{pmatrix},
&
\SWAP &:= 
 \begin{pmatrix}
    1 & 0 & 0 & 0\\
    0 & 0 & 1 & 0\\
    0 & 1 & 0 & 0\\
    0 & 0 & 0 & 1
 \end{pmatrix}.
\end{align}

Each gate may be depicted as a \emph{circuit element} in which qubits are
represented by horizontal lines (\emph{wires}) entering on the left and exiting on the
right.  The circuit elements depicting the gates above are
\begin{align*}
&\Qcircuit{
 & \targ & \qw 
}
&&
\Qcircuit{
 & \gate{H} & \qw 
}
\\[3ex]
&\Qcircuit{
 & \gate{S} & \qw 
}
&&
\Qcircuit{
 & \qswap & \qw \\
 & \qswap \qwx & \qw
}
\end{align*}

The \emph{controlled-$U$ gate}, denoted $\Lambda(U)$, applies the gate $U$
to a \emph{target} state conditioned on a \emph{control} qubit being in the
state $|1\>$.  The matrix and circuit element representing $\Lambda(U)$ are

\begin{align*}
\Lambda(U) &:= 
 \begin{pmatrix}
    I & 0 \\
    0 & U 
 \end{pmatrix},
&
\raisebox{3ex}{
\Qcircuit{
 & \ctrl{1} & \qw \\
 & \gate{U} & \qw 
}
}
\end{align*}
where $I$ represents the identity matrix with $\dim I = \dim U$.

We use two non-standard shorthand notations for gates in this paper.  The
first is that we denote gates controlled by the state $|0\>$ rather than the
state $|1\>$ as
\begin{align*}
\Qcircuit{
 & \ctrlo{1} & \qw \\
 & \gate{U} & \qw 
}
&\raisebox{-1.6em}{\quad $:=$\quad } 
\Qcircuit @C=1em {
 & \targ & \ctrl{1} & \targ & \qw \\
 & \qw   & \gate{U} & \qw   & \qw 
}
\end{align*}
The second is that we denote a cascaded array of nearest-neighbor swap gates
as
\begin{align*}
\Qcircuit @C=1em @R=0em {
  & \ajlmultigate{5}{\Sigma}{1em} & \qw \\
  & \ajlghost{\Sigma}{1em}        & \qw \\
  & \ajlghost{\Sigma}{1em}        & \qw \\
  & \ajlghost{\Sigma}{2.29em}     & \qw \\
  & \ajlghost{\Sigma}{1em}        & \qw \\
  & \ajlghost{\Sigma}{1em} & \qw
}
&\raisebox{-3.6em}{\ $=$\ }
\Qcircuit @C=1em @R=1em {
 & \qswap      & \qw         & \qw             & \qw           & \qw
& \qw         & \qw \\
 & \qswap \qwx & \qswap      & \qw             & \qw           & \qw
& \qw         & \qw \\
 & \qw         & \qswap \qwx & \qswap \qwx[1]  & \qw           & \qw
& \qw         & \qw \\
 &             &             &                 & \push{\ddots} &
&             &     \\
 & \qw         & \qw         & \qw             & \qw           & \qswap \qwx
& \qswap      & \qw \\
 & \qw         & \qw         & \qw             & \qw           & \qw
& \qswap \qwx & \qw \\
}
\end{align*}

\begin{figure}[ht]
\centerline{
\Qcircuit @C=1em @R=1em {
\lstick{\raisebox{-3em}{$|\textrm{Data}\>$~~}}    & \qw       & \gate{S}  & \qw \\
                                                  & \gate{H}  & \ctrl{-1} & \qw \\ 
                                                  &           &           &     \\
                                                  & \ctrl{-2} & \qw       & \qw \\
\lstick{\raisebox{1em}{$|\textrm{Program}\>$~~}}  & \qw       & \ctrl{-4} & \qw%
\gategroup{1}{1}{2}{1}{1em}{\{} \gategroup{4}{1}{5}{1}{1em}{\{}
}
}
\caption{A simple programmable circuit over the Kitaev gate basis.}
\label{fig:programmable-circuit}
\end{figure}
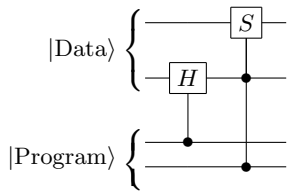

A \emph{gate basis} is a set of gates; a circuit is expressed \emph{over} a
gate basis $\calG$ if all of its gates are elements of $\calG$.  A circuit
family is \emph{uniformly generated} if a description of its elements can be
constructed by finite means, \eg, by a Turing machine \cite{Turing:1937a}.
A gate basis $\calG$ is \emph{universal} if for any unitary transformation
$U$ and for any desired precision $\epsilon$, there is a uniformly generated
circuit family over $\calG$ that can approximate $U$ to within $\epsilon$.
An example of a universal gate basis is the \emph{Kitaev gate basis} $\{H,\
\Lambda(S)\}$ \cite{Kitaev:1997b}.  A \emph{programmable} circuit is one
that can be conceptually divided into ``program'' and ``data'' regions,
where the state of the program region controls which gates are to be applied
to the state of the data region.  For example, Fig.\
\ref{fig:programmable-circuit} depicts a programmable circuit over the gate
basis $\{\Lambda(H),\ \Lambda^2(S)\}$ in which the program controls which
gates from the Kitaev basis are to be applied to the data.  A gate basis
$\calG$ is \emph{programmably universal} if for any desired precision
$\epsilon$ and any number of qubits $n$, there is a uniformly generated
programmable circuit $V$ over $\calG$ such that for any $n$-qubit unitary
transformation $U$, there is a uniformly generated program state that
enables $V$ to approximate $U$ to within $\epsilon$ on the data portion of
$V$.

%

\subsection{Programmably simulating circuits over the Kitaev gate basis
exactly}

\begin{figure}[ht] 
\includegraphics[height=1.5in]{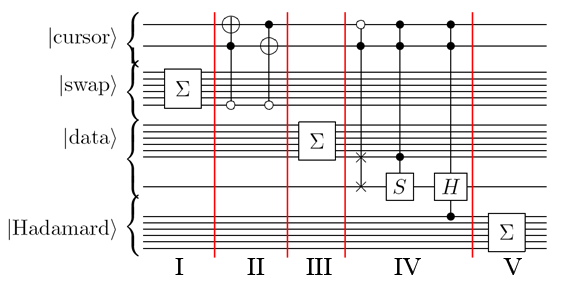}
\caption{The quantum circuit $V$, whose action is described in the text.}
\label{fig:V-circuit}
\end{figure}

Consider the programmable circuit $V$ depicted in Fig.~\ref{fig:V-circuit}.
Its program region is decomposed into three parts labeled \emph{cursor},
\emph{swap}, and \emph{Hadamard}.  Given a description of a quantum circuit
$U$ on $n$ qubits containg $T$ gates that is expressed over the Kitaev gate
basis, the program region of $V$ can be initialized so that $\bigO(nT)$
iterations of $V$ will simulate $U$ exactly.  To see how, we first trace
through one pass of the circuit $V$ to see how it responds to various
initializations of the program regions.  We then describe how to initialize
these regions to carry out the simulation of $U$.

\vspace{1ex}

\noindent\textbf{Phase I}.  The $\Sigma$ gate moves the upper-most
\emph{swap} qubit state to the lower-most \emph{swap} qubit and moves every
other \emph{swap} qubit state up one qubit to accommodate.

\vspace{1ex}

\noindent\textbf{Phase II}.  If the lower-most \emph{swap} qubit is in the
state $|0\>$, then the two-qubit \emph{cursor} state permutes from $|00\>
\to |00\>$ and from $|01\> \to |10\> \to |11\> \to |01\>$. We depict this
permutation in glyphs as $\nothing \to \nothing$ and $\cycle \to \holdcycle
\to \gatecursor \to \cycle$.  The reason for this choice of glyphs will
become apparent later.

\vspace{1ex}

\noindent\textbf{Phase III}.  The $\Sigma$ gate moves the upper-most
\emph{data} qubit state to the penultimate \emph{data} qubit and moves the
state of every \emph{data} qubit in between up one to accommodate.

\vspace{1ex}

\noindent\textbf{Phase IV}.  If the two-qubit \emph{cursor} is in the state
$\cycle$, then the penultimate \emph{data} qubit state is swapped with the
final \emph{data} qubit state.  If the \emph{cursor} is in the state
$\gatecursor$, then first $\Lambda(S)$ is applied between the last two
\emph{data} qubits, followed by an application of $H$ to the last
\emph{data} qubit, conditioned on the upper-most \emph{Hadamard} qubit being
in the state $|1\>$.  If the \emph{cursor} is in any other state, no action
is performed on the \emph{data} qubits.

\vspace{1ex}

\noindent\textbf{Phase V}.  The $\Sigma$ gate moves the upper-most
\emph{Hadamard} qubit state to the lower-most \emph{Hadamard} qubit and
moves every other \emph{Hadamard} qubit state up one to accommodate.

\vspace{2ex}

Given this action of one iteration of $V$, we program $V$ so that it follows
a sequence of ``behaviors'' that depend primarily on the state of the
two-qubit \emph{cursor}:

\vspace{1ex}

\noindent\textbf{Behavior 1.}  The two-qubit \emph{cursor} is initialized in
the state $\cycle$ and the first $n-k$ \emph{swap} qubits are initialized in
the state $|1\>$.  The effect of applying $V$ a total of $n-k$ times given
this initialization is to cyclically permute the states of qubits in each
program region and in the data region upwards $n-k$ times.  Importantly, the
state of the $k$th data qubit from the top before this behavior is stored in
the lower-most data qubit after this behavior.  The glyph $\cycle$ is
intended to remind that when the cursor is in this state, the \emph{data}
qubit states are all cyclically permuted.

\vspace{1ex}

\noindent\textbf{Behavior 2.}  The \emph{cursor} is initialized in
the state $\cycle$, the upper-most \emph{swap} qubit state is initialized in
the state $|0\>$ and the next $n-j-2$ \emph{swap} qubits are initialized in
the state $|1\>$.  The effect of applying $V$ once given this initialization
is to cyclically permute the \emph{swap} qubit's states upwards once, cycle
the \emph{cursor} state from $\cycle$ to $\holdcycle$, cyclically permute
all but the state of the last \emph{data} qubit upwards once, and cyclically
permute all \emph{Hadamard} qubit states upwards once.  The effect of
applying $V$ a total of $n-j-2$ more times given this initialization is to
repeat this behavior $n-j-2$ times, except it will not cycle the state of
the \emph{cursor} qubits.  Importantly, the state of the $j$th \emph{data}
qubit from the top before this behavior becomes stored in the third-to-last
\emph{data} qubit after this behavior.  The glyph $\holdcycle$ is intended
to remind that when the cursor is in this state, the effect is to cyclically
permute all but the last \emph{data} qubit state.

\vspace{1ex}

\noindent\textbf{Behavior 3.} The \emph{cursor} is initialized in
the state $\holdcycle$ and the upper-most \emph{swap} qubit is initialized
in the state $|0\>$. The effect of applying $V$ once given this
initialization is to cyclically permute the \emph{swap} qubit states upwards
once, cycle the \emph{cursor} state from $\holdcycle$ to $\gatecursor$,
cyclically permute all but the last \emph{data} qubit state upwards once,
apply $\Lambda(S)$ between the last two \emph{data} qubits, apply $H$ to the
last \emph{data} qubit if the upper-most \emph{Hadamard} qubit is in the
state $|1\>$, and cyclically permute the \emph{Hadamard} qubit states
upwards once.  Importantly, the states of qubits $k$ and $j$ manipulated in
behaviors $1$ and $2$ have the gate $\Lambda(S)$ applied between them and
the state of qubit $k$ has the gate $H$ applied to it conditioned on the
state of the upper-most \emph{Hadamard} qubit before this behavior.  The
glyph $\gatecursor$ is intended to remind that when the cursor is in this
state, the effect is to apply gates from the Kitaev gate basis to the
\emph{data} qubits.

\vspace{1ex}

\noindent\textbf{Behavior 4.} The \emph{cursor} is initialized in
the state $\gatecursor$ and the upper-most \emph{swap} qubit is initialized
to the state $|0\>$.  The effect of applying $V$ once given this
initialization is to cyclically permute the states of the \emph{swap} qubits
upwards once, cycle the \emph{cursor} state from $\gatecursor$ to $\cycle$,
and cyclically permute the \emph{data} and \emph{Hadamard} qubits upwards
once.  Importantly, after this behavior, the \emph{cursor} is in the same
state it started in for behavior 1.

Behaviors $1$--$3$ describe how, given a suitable initialization of program
qubits, the gate $(I\otimes H)\Lambda(S)$ or $\Lambda(S)$ can be applied to
any two desired data qubits.  Moreover, behavior $4$ shows that by
lengthening the program region, any sequence of such gates can be
implemented.  However, the Kitaev gate basis is not $\{\Lambda(S), (I\otimes
H)\Lambda(S)\}$, but $\{H, \Lambda(S)\}$.  In order to apply $H$ to any
desired qubit, we use a trick: we program $V$ to apply $(I\otimes
H)\Lambda(S) \cdot \Lambda(S)^3$.  Because $\Lambda(S)^4 = I$, this is the
same as applying $H$.

If a circuit $U$ over the Kitaev gate basis contains
$T_{\Lambda(S)}$ controlled-$S$ gates and $T_H$ Hadamard gates on $n$ 
data qubits, then the number of program region qubits (and gates) needed by
$V$ is at most $4n(T_{\Lambda(S)} + 2T_H)$, so the simulation by $V$ is
\emph{linear} in the size of $U$.

It is worth noting that $V$ is defined over a gate basis that contains gates
that are not spatially local when the qubits are arranged in a line.
Spatial locality in one dimension could have been achieved much more easily
(and without the need for programmability) with, say, the gate basis $\{H,
\Lambda(S), \SWAP\}$.  Nevertheless, all the nonlocality in $V$ is between
the cursor program qubits and the other data qubits and enables $V$ to be
repeated in a simple fashion to simulate arbitrary circuits $U$ over the
Kitaev gate basis exactly.  By replacing each of the non-cursor qubits with
eight-state systems (to simulate one data qubit plus two cursor qubits), it
is possible to derive a related programmable circuit that requires only
spatially local gates in one dimension.  Instead of going through this
exercise, we move on to how to construct a Hamiltonian that would simulate
such a circuit via a continuous-time quantum walk.

%

\section{A family of nearest-neighbor Hamiltonians coupling 8-level systems
on a ring}
\label{sec:architecture}

In this section, we construct a Hamiltonian $H_8$ acting on nearest-neighbor
eight-state systems on a ring.  The interactions in $H_8$ are closely
related to the gates used by the circuit $V$ of Fig.~\ref{fig:V-circuit} of
the last section.  We defer a discussion of how $H_8$ is used to achieve
universal quantum computation to subsequent sections.

%

\subsection{State space and geometry}

Let $V$ be the circuit depicted in Fig.\ \ref{fig:V-circuit}, and let
$V^{(K)}$, $K = \bigO(nT)$, be a circuit with suitably initialized inputs
that simulates a $T$-gate, $n$-qubit circuit over the Kitaev gate basis.
Let $\calH_8$ denote the Hilbert space of a quantum system having eight
possible orthogonal states and let $\calH_8^{\otimes K}$ denote the Hilbert
space of a collection of $K$ of these systems arranged in a one-dimensional
ring.  For convenience, we will think of the eight-state system as arising
from the combined state space of quantum systems having two and four states
respectively.  In other words, we will use the conceptual decomposition of
the Hilbert space of each of these systems as $\calH_8 =
\calH_2\otimes\calH_4$, where $\calH_d$ indicates a Hilbert space of
dimension $d$.  
\begin{figure}[b]
\begin{tabular}{lc@{}c@{}l@{}}
& \emph{swap} & \emph{data} & \emph{Hadamard}\\
 & \nothing \nothing \nothing \nothing \nothing \nothing \gatecursor\firstboundary & \nothing \nothing \nothing \nothing \secondboundary & \nothing \nothing \nothing \nothing \nothing \thirdboundary\\
 & \zero \one \zero \zero \one \one \one \firstboundary & \zeroone \zeroone \zeroone \zeroone \secondboundary & \zero \one \one \zero \zero \thirdboundary\\
\vspace{0.5em}
\end{tabular}

\includegraphics[width=0.4\textwidth]{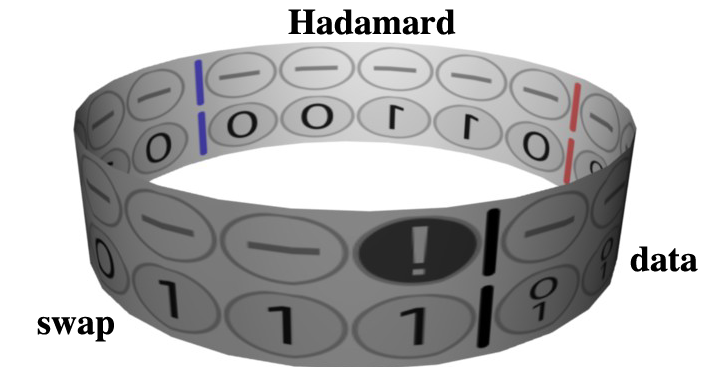}
\caption{Architecture Layout (Color online)}
\label{fig:layout}
\end{figure}

As illustrated in Figure \ref{fig:layout}, this ring can be divided into two
parallel ``lines'' and three regions.  This layout mimics that of the
circuit $V$ rotated on its side, but with periodic boundary conditions.
However instead of having a cursor program region, each qubit on the lower
line has its own adjacent $4$-state cursor system on the upper line that can
be in one of the ``active'' states $\cycle$, $\holdcycle$, or $\gatecursor$,
or in $\nothing$, a new ``inactive'' state not used by $V$.  Just as the
program region qubits in $V$ are always in classical states $|0\>$ or $|1\>$
for any simulation of a quantum circuit, so too will the qubits in the
program regions of this ring always be in classical states, which we label
schematically by the glyphs $\zero$ and $\one$.  However, qubits in the data
region may be in any superposition of $|0\>$ and $|1\>$, and indeed even
entangled with the state of all other data qubits, so we represent an
arbitrary state of a data qubit by the glyph $\zeroone$.  Table
\ref{fig:symbols} is a legend for the glyphs we use to denote the cursor
states, qubit states, and region boundaries in this geometry.

\begin{figure}
\begin{center}
\begin{tabular}{|l|l|}
\hline Cursor Line  & Program \& Data Line \\
\hline \nothing: Inactive/Non-cursor site & \zero: ``Classical'' zero bit\\
\gatecursor: Gate phase cursor & \one: ``Classical'' one bit\\
\cycle: Cycle phase cursor & \zeroone: Arbitrary qubit state\\
\holdcycle: Hold-cycle phase cursor &  \\
\arbitraryqudit: Any one of $\cycle$,$\holdcycle$,$\gatecursor$ &\\
\hline \multicolumn{2}{|c|}{Boundary Markers}\\
\hline
\multicolumn{2}{|l|}{\hspace{1em}\firstboundary\hspace{1em} Between \emph{swap} and \emph{data} }\\
\multicolumn{2}{|l|}{\hspace{1em}\secondboundary\hspace{1em} Between \emph{data} and \emph{Hadamard} }\\
\multicolumn{2}{|l|}{\hspace{1em}\thirdboundary\hspace{1em} Between \emph{Hadamard} and \emph{swap} }\\
\hline
\end{tabular}
\caption{Glyphs for cursor states, qubits states, and region boundaries.  (Color online.)}
\label{fig:symbols}
\end{center}
\end{figure}

%

\subsection{Hamiltonian construction}

Like Feynman's Hamiltonian (\ref{eq:H_feynman}), we construct $H_8$ on
$\calH_8$ as a sum of terms with $g_i$ representing forward computation and
$g_i^\dagger$ representing backwards computation:
\begin{align}
\label{eq:gHamiltonian}
H &:= \sum_{i=1}^K g\ssc{i} + g_i^\dagger.
\end{align}

Each operator $g_i$ is an interaction involving only the 8-state systems at
sites $i$ and $(i+1) \bmod K$ in the ring.  A given $g_i$ is generically
responsible for two things.  Firstly, it examines the current cursor state
at site $i$, replaces it with $\nothing$ and appropriately sets the cursor
at site $i+1$.  This effectively passes the cursor down the ring and ensures
that only a single site is ``active'' (non-$\nothing$).  Secondly, a $g_i$
is charged with performing any necessary gates on the qubits at sites $i$
and $i+1$, which depend on both the location $i$ and the current state of
the cursor.  

At most of the ring sites, $g_i$ merely swaps the state of the cursor.
Namely,
\begin{align}
\label{eq:general-swap-rule}
g_i &:= 
\ket{\begin{array}{c}
\nothing \, \arbitraryqudit\\
\zero \, \arbitraryqubit
\end{array}}\!\!\bra{\begin{array}{c}
\arbitraryqudit \, \nothing\\
\arbitraryqubit \, \zero
\end{array}} 
\ +\  \ket{\begin{array}{c}
\nothing \, \arbitraryqudit\\
\one \, \arbitraryqubit 
\end{array}}\!\!\bra{\begin{array}{c}
\arbitraryqudit \, \nothing\\
\arbitraryqubit \, \one 
\end{array}},
\end{align}
where $\arbitraryqudit$ represents an unspecified \cycle,\holdcycle\ or
$\gatecursor$ state that is preserved during the transition and
\arbitraryqubit\ represents either a \zero\  or \one\ that is preserved
during the transition.  In other words, Eq.\ (\ref{eq:general-swap-rule}) is
actually a sum of terms, one for each possible value of $\arbitraryqubit$
and $\arbitraryqudit$.

The only places where $g_i$ deviates from this behavior is near the
boundaries between regions.  This is because the circuit $V^{(K)}$ consists
of $\SWAP$ gates everywhere except near these boundaries.  We define how $g_i$
acts near these boundaries below:

\begin{itemize}

\item{}\emph{Swap Region Near the Swap-Data Boundary}

In order to simulate the action of $V^{(K)}$, which has a definite beginning
and end, by $H_8$, which has periodic boundary conditions, we replace Eq.\
(\ref{eq:general-swap-rule}) for the last two states in the swap program
region with projections onto \emph{start} and \emph{stop}
states:
\begin{align}
\label{eq:start-stop-projections}
g_i &:=
\frac{1}{2}
\ket{\begin{array}{c}
\nothing \, \gatecursor \firstboundary \\
\one \, \one \firstboundary 
\end{array}}\!\!\bra{\begin{array}{c}
\nothing \, \gatecursor \firstboundary \\
\one \, \one \firstboundary
\end{array}}
\ +\ 
\frac{1}{2}
\ket{\begin{array}{c}
\gatecursor \, \nothing \firstboundary \\
\one \, \zero \firstboundary 
\end{array}}\!\!\bra{\begin{array}{c}
\gatecursor \, \nothing \firstboundary \\
\one \, \zero \firstboundary
\end{array}},
\end{align}
where the factors of $1/2$ are to accommodate the fact that both $g_i$ and
$g_i^\dagger$ are included in $H_8$.

The second term in (\ref{eq:start-stop-projections}) is a projection onto a
stop state that is never reached by a valid programming of $V^{(K)}$.
Specifically, ``Behavior 4'' in Sec.\ \ref{sec:V-circuit} prevents this from
happening.  (A cursor in the state $\gatecursor$ never sees a swap program
qubit in the state $\one$.)  By augmenting the swap program with one extra
qubit in the state $\one$, the general swapping interaction
(\ref{eq:general-swap-rule}) will evolve a state with a $\gatecursor$ atop
$\one$ to the last two swap program sites discussed here and stop
propagating further.  Of course, because a $g_i^\dagger$ term is in $H_8$
for every $g_i$ term in $H_8$, it is possible for evolution to ``undo'' some
of the computation and compute in reverse from the stop state.  However, no
additional forward computation will accrue after this stop state is reached.

The first term in (\ref{eq:start-stop-projections}) is also one that is
never reached by a valid programming of $V^{(K)}$.  Moreover, it is never
reached by the extended ``stop state'' implementation by $H_8$ just
discussed.  Indeed, it is precisely the stop state projection term that
prevents the start state from ever being reached.  Hence no computation can
be ``undone'' by $g_i^\dagger$ terms once this start state is reached (in
reverse).

The net effect of incorporating stop and start state interactions is that
evolution by $H_8$ on a ring can be ``unfurled'' into an equivalent
interaction by a related Hamiltonian on a line.  This will be described
in more detail in Sec.\ \ref{sec:quantum-walk-universality}.

To simulate the action of $V^{(K)}$ on these two qubits for a \emph{valid}
programming of the swap program register, we need to add to $g_i$ of Eq.\
(\ref{eq:start-stop-projections}) the following terms:
\begin{align}
g_i &:=
\ket{\begin{array}{c}
\nothing \, \holdcycle \firstboundary \\
\zero \, \arbitraryqubit \firstboundary 
\end{array}}\!\!\bra{\begin{array}{c}
\holdcycle \, \nothing \firstboundary \\
\arbitraryqubit \, \zero \firstboundary
\end{array}}
\ +\ 
\ket{\begin{array}{c}
\nothing \, \holdcycle \firstboundary \\
\one \, \arbitraryqubit \firstboundary 
\end{array}}\!\!\bra{\begin{array}{c}
\holdcycle \, \nothing \firstboundary \\
\arbitraryqubit \, \one \firstboundary
\end{array}}
\\[1em] \nonumber
&+\ 
\ket{\begin{array}{c}
\nothing \, \cycle \firstboundary \\
\zero \, \arbitraryqubit \firstboundary 
\end{array}}\!\!\bra{\begin{array}{c}
\cycle \, \nothing \firstboundary \\
\arbitraryqubit \, \zero \firstboundary
\end{array}}
\ +\ 
\ket{\begin{array}{c}
\nothing \, \cycle \firstboundary \\
\one \, \arbitraryqubit \firstboundary 
\end{array}}\!\!\bra{\begin{array}{c}
\cycle \, \nothing \firstboundary \\
\arbitraryqubit \, \one \firstboundary
\end{array}}
\\[1em] \nonumber
&+\ 
\ket{\begin{array}{c}
\nothing \, \gatecursor \firstboundary \\
\arbitraryqubit \, \zero \firstboundary 
\end{array}}\!\!\bra{\begin{array}{c}
\gatecursor \, \nothing \firstboundary \\
\zero \, \arbitraryqubit \firstboundary
\end{array}}.
\end{align}

Finally, although the following state only arises when $V$ simulates one of
the rather banal one-gate circuits $U = (I\otimes H)\Lambda(S)$ or $U =
\Lambda(S)$, we add to $g_i$ a projection onto a second stop state for
completeness' sake:
\begin{align}
g_i &:=
\frac{1}{2}
\ket{\begin{array}{c}
\gatecursor \, \nothing \firstboundary \\
\one \, \one \firstboundary 
\end{array}}\!\!\bra{\begin{array}{c}
\gatecursor \, \nothing \firstboundary \\
\one \, \one \firstboundary
\end{array}}
\end{align}
%

\item{}\emph{Swap-Data Boundary}

To simulate the permutation of cursor states $\cycle \to \holdcycle \to
\gatecursor \to \cycle$ enacted by $V^{(K)}$ when the state of the lower-most
qubit in the swap program is a $0$, we replace
(\ref{eq:general-swap-rule}) at the swap-data boundary with
\begin{align}
g_i &:=
\ket{\begin{array}{c}
\nothing \firstboundary \holdcycle\\
\zero \firstboundary \arbitraryqubit 
\end{array}}\!\!\bra{\begin{array}{c}
\cycle \firstboundary \nothing\\
\zero \firstboundary \arbitraryqubit 
\end{array}} 
\ +\ 
\ket{\begin{array}{c}
\nothing \firstboundary \gatecursor\\
\zero \firstboundary \arbitraryqubit 
\end{array}}\!\!\bra{\begin{array}{c}
\holdcycle \firstboundary \nothing\\
\zero \firstboundary \arbitraryqubit 
\end{array}} 
\\[1ex] \nonumber
&+\ 
\ket{\begin{array}{c}
\nothing \firstboundary \cycle\\
\zero \firstboundary \arbitraryqubit 
\end{array}}\!\!\bra{\begin{array}{c}
\gatecursor \firstboundary \nothing\\
\zero \firstboundary \arbitraryqubit 
\end{array}}.
\end{align}

To simulate the movement of the cursor from the swap region to the data
region when the last qubit's state of the swap program is a $1$ (and no
cycling of the cursor's state is performed), we add to $g_i$ at this
location one more term:
\begin{align}
\label{eq:swap-data-boundary-rule}
g_i :=
\ket{\begin{array}{c}
\nothing \firstboundary \arbitraryqudit\\
\one \firstboundary \arbitraryqubit 
\end{array}}\!\!\bra{\begin{array}{c}
\arbitraryqudit \firstboundary \nothing\\
\one \firstboundary \arbitraryqubit 
\end{array}}.
\end{align}


\item{}\emph{Data Region Near the Data-Hadamard Boundary}

Because $V^{(K)}$ swaps the last two data qubits only when the cursor is in
the state $\cycle$, and because $V^{(K)}$ only applies a $\Lambda(S)$
between these qubits when the cursor is in the state $\gatecursor$, we
replace (\ref{eq:general-swap-rule}) at the last two data sites with
\begin{align}
g_i &:=
\ket{\begin{array}{c}
\nothing \, \holdcycle \secondboundary \\
\arbitraryqubit \, \zero \secondboundary 
\end{array}}\!\!\bra{\begin{array}{c}
\holdcycle \, \nothing \secondboundary \\
\arbitraryqubit \, \zero \secondboundary
\end{array}}
\ +\ 
\ket{\begin{array}{c}
\nothing \, \holdcycle \secondboundary \\
\arbitraryqubit \, \one \secondboundary 
\end{array}}\!\!\bra{\begin{array}{c}
\holdcycle \, \nothing \secondboundary \\
\arbitraryqubit \, \one \secondboundary
\end{array}}
\\[1em] \nonumber
&+\ 
\ket{\begin{array}{c}
\nothing \, \cycle \secondboundary \\
\zero \, \arbitraryqubit \secondboundary 
\end{array}}\!\!\bra{\begin{array}{c}
\cycle \, \nothing \secondboundary \\
\arbitraryqubit \, \zero \secondboundary
\end{array}}
\ +\ 
\ket{\begin{array}{c}
\nothing \, \cycle \secondboundary \\
\one \, \arbitraryqubit \secondboundary 
\end{array}}\!\!\bra{\begin{array}{c}
\cycle \, \nothing \secondboundary \\
\arbitraryqubit \, \one \secondboundary
\end{array}}
\\[1em]
&+\ 
\ket{\begin{array}{c}
\nothing \, \gatecursor \secondboundary \\
\overline{\arbitraryqubit} \, \overline{\zero} \secondboundary 
\end{array}}\!\!\bra{\begin{array}{c}
\gatecursor \, \nothing \secondboundary \\
\arbitraryqubit \, \zero \secondboundary
\end{array}}
\ +\ 
\ket{\begin{array}{c}
\nothing \, \gatecursor \secondboundary \\
\overline{\arbitraryqubit} \, \overline{\one} \secondboundary 
\end{array}}\!\!\bra{\begin{array}{c}
\gatecursor \, \nothing \secondboundary \\
\arbitraryqubit \, \one \secondboundary
\end{array}}, \nonumber
\end{align}
where the bar over the qubits in the last two terms indicates that a
$\Lambda(S)$ gate has been applied between them.

\item{}\emph{Data-Hadamard Boundary}

Because no swapping occurs across the data-Hadamard boundary and because
additionally a Hadamard gate is applied to the state of the last data qubit
if the state of the upper-most Hadamard program qubit is a $1$ and the
cursor is in the state $\gatecursor$, we replace Eq.\
(\ref{eq:general-swap-rule}) at the data-Hadamard boundary with
\begin{align}
g_i &:=
\ket{\begin{array}{c}
\nothing \secondboundary \holdcycle \\
\arbitraryqubit \secondboundary \zero
\end{array}}\!\!\bra{\begin{array}{c}
\holdcycle \secondboundary \nothing \\
\arbitraryqubit \secondboundary \zero 
\end{array}}
\ +\ 
\ket{\begin{array}{c}
\nothing \secondboundary \holdcycle \\
\arbitraryqubit \secondboundary \one 
\end{array}}\!\!\bra{\begin{array}{c}
\holdcycle \secondboundary \nothing \\
\arbitraryqubit \secondboundary \one 
\end{array}}
\\[1em] \nonumber
&+\ 
\ket{\begin{array}{c}
\nothing \secondboundary \cycle \\
\arbitraryqubit \secondboundary \zero 
\end{array}}\!\!\bra{\begin{array}{c}
\cycle \secondboundary \nothing \\
\arbitraryqubit \secondboundary \zero 
\end{array}}
\ +\ 
\ket{\begin{array}{c}
\nothing \secondboundary \cycle \\
\arbitraryqubit \secondboundary \one 
\end{array}}\!\!\bra{\begin{array}{c}
\cycle \secondboundary \nothing \\
\arbitraryqubit \secondboundary \one 
\end{array}}
\\[1em]
&+\ 
\ket{\begin{array}{c}
\nothing \secondboundary \gatecursor \\
\arbitraryqubit \secondboundary \zero
\end{array}}\!\!\bra{\begin{array}{c}
\gatecursor \secondboundary \nothing \\
\arbitraryqubit \secondboundary \zero 
\end{array}}
\ +\ 
\ket{\begin{array}{c}
\nothing \secondboundary \gatecursor \\
\overline{\arbitraryqubit} \secondboundary \one 
\end{array}}\!\!\bra{\begin{array}{c}
\gatecursor \secondboundary \nothing \\
\arbitraryqubit \secondboundary \one 
\end{array}}, \nonumber
\end{align}
where the bar over the qubit in the last term indicates that a
Hadamard gate has been applied to it.

\item{}\emph{Hadamard-Swap Boundary}

The only subtlety about the Hadamard-swap boundary is that qubit states are
not swapped across them in $V^{(K)}$.  Hence Eq.\
(\ref{eq:general-swap-rule}) is replaced at this boundary by
\begin{align}
g_i :=
\ket{\begin{array}{c}
\nothing \thirdboundary \arbitraryqudit\\
\one \thirdboundary \arbitraryqubit 
\end{array}}\!\!\bra{\begin{array}{c}
\arbitraryqudit \thirdboundary \nothing\\
\one \thirdboundary \arbitraryqubit 
\end{array}}
\ +\ 
\ket{\begin{array}{c}
\nothing \thirdboundary \arbitraryqudit\\
\zero \thirdboundary \arbitraryqubit 
\end{array}}\!\!\bra{\begin{array}{c}
\arbitraryqudit \thirdboundary \nothing\\
\zero \thirdboundary \arbitraryqubit 
\end{array}}.
\end{align}

\end{itemize}

Given this definition for our Hamiltonian $H_8$, our next task is to
demonstrate how it may be programmed to simulate quantum circuits.  In the
next section, we show how this can be done when $H_8$ drives a
continuous-time quantum walk.  In the subsequent section, we show how this
can be done when (a slightly modified version of) $H_8$ is the final
Hamiltonian of an adiabatic algorithm.

%

\section{Universal quantum computation via a continuous-time quantum walk
driven by $H_8$}
\label{sec:quantum-walk-universality}

The Hamiltonian $H_8$ defined in the previous section enables universal
quantum computation by driving a continuous-time quantum walk in the
following way.  First choose a precision $\epsilon$ and a quantum circuit
$W$ to simulate to within precision $\epsilon$.  Then, construct a circuit
$U$ over the Kitaev gate basis that approximates $W$ to within $\epsilon$.
This can be done using standard techniques, \eg, by the Solovay-Kitaev
algorithm \cite{Dawson:2005a}.  Next, determine how to program the circuit
$V$ defined in Sec.\ \ref{sec:V-circuit} so that it simulates $U$ exactly.
Initialize the \emph{swap}, \emph{data}, and \emph{Hadamard} regions of
$\calH_4\otimes \calH_2$ to the same states that one would initialize the
\emph{swap}, \emph{data}, and \emph{Hadamard} regions for $V$.  However,
because the \emph{swap} region on the 1D ring has one more state than in the
$V$ circuit, initialize the last \emph{swap} qubit on the 1D ring to be in
the state $|1\>$.  (Note that a valid programming obtained from $V$ will
necessarily also have the penultimate \emph{swap} qubit also initialized to
the state $|1\>$.)  Finally, initialize the cursor line on the 1D ring to be
$\nothing$ everywhere, except over the last \emph{swap} qubit, where it is
in the state $\gatecursor$.  We will denote the state of the 1D ring so
initialized by $|\psi_0\>$.

Only two $g_i$ terms in $H_8$ act nontrivially on $|\psi_0\>$.  The first is
the \emph{start} state projector (\ref{eq:start-stop-projections}) and the
second is the transition term across the \emph{swap}-\emph{data} boundary
(\ref{eq:swap-data-boundary-rule}).  Thus evolution by $H_8$ can either keep
the ring in the state $|\psi_0\>$ or advance it with some amplitude to a
unique successor state $|\psi_1\>$.  If the ring is in the state
$|\psi_1\>$, the Hamiltonian $H_8$ only couples it back to $|\psi_0\>$ or to
a unique successor state $|\psi_2\>$ given by the general swap rule
(\ref{eq:general-swap-rule}).  This line of reasoning continues, with a
unique successor state and predecessor state existing for $|\psi_t\>$ in the
ring until the final $t = \overline{T} := \bigO(nT)$.  In this
configuration, only the \emph{stop} state projector and the predecessor
state match.  The Hamiltonian $H_8$ can therefore be restricted to a
subspace of size $\overline{T} + 1$, on which it looks like
\begin{align} \label{eq:HeffectiveWalk}
H_8^{(\text{eff})} &:=
\sum_{i=1}^{\overline{T}} |\psi_t\>\<\psi_{t-1}| + |\psi_{t-1}\>\<\psi_t| \\
&+ |\psi_0\>\<\psi_0| +
 |\psi_{\overline{T}}\>\<\psi_{\overline{T}}|.
\end{align}

At this point, one can argue as Feynman originally did \cite{Feynman:1986a}
that evolution by $H_8$ is the same as by a quantum walk on a line.
Straightforward analysis demonstrates that the time at which this walk has a
maximum amplitude for moving from $|\psi_0\>$ to $|\psi_{\overline{T}}\>$ is
time $T/2$, at which the amplitude is
\begin{align}
\<\psi_{\overline{T}}|e^{-iH_8T/2}|\psi_0\> \approx {\overline{T}}^{-1/3}.
\end{align}
A detailed analysis in terms of Bessel functions of why this is the case can
be found in numerous places, for example in \cite{Childs:2003a}.

Given these insights, a continuous-time quantum walk driven by $H_8$ can be
made to simulate $V$ arbitrarily well in several different ways:

\begin{itemize}

\item{}

One could simply repeat the preparation and evolution
$\bigO(\overline{T}^{2/3})$ times, availing upon the Chernoff bound that it
is exponentially likely in $\overline{T}$ that one of the final measurements will
find the system in the state $|\psi_{\overline{T}}\>$, representing the output of
the computation.

\item{}

One could replace the projection onto the stop state in $H_8$ with
$\bigO(\overline{T}^{2/3})$ general swap transitions defined by Eq.\
(\ref{eq:general-swap-rule}) on $\bigO(\overline{T}^{2/3})$ additional qubits in
the swap program region.  By use of the Chernoff bound once again, it
becomes exponentially likely in $\overline{T}$ that the cursor is in one of these
new ``dummy'' locations.  Note that simply adding $\bigO(\overline{T}^{2/3})$
identity gates to the end of the circuit $V$ being simulated is not trivial
as it was in similar refs.\ \cite{Feynman:1986a, FOCS:2004a} because $V$
simulates the identity by applying $\Lambda(S)^4$.

\item{}

One could replace both the start state and stop state projections in $H_8$
by additional ``runway'' and ``landing pad'' qubit chains of size
$\bigO(\overline{T}^{2/3})$ that are swapped by the general swap rule
(\ref{eq:general-swap-rule}).  As shown by Feynman \cite{Feynman:1986a} (and
argued in greater detail in \cite{Osborne:2004a}) this will allow the
computation to proceed ballistically with arbitrarily high probability.

\item{}

One could redefine $H_8$ so that the $t$th term had a coefficient
$\sqrt{\overline{T}(\overline{T}-t)}$, enabling perfect fidelity state transfer to the
final state of the computation \cite{Christandl:2004a,Christandl:2005a}.  

\end{itemize}

Running a quantum algorithm in this model is particularly simple.  One first
initializes the state $|\psi_0\>$, then waits a time $T/2$, and finally
measures the state of the ring destructively using one of the methods
described above.  Because no dynamical controls are required during the
course of the computation, ``gate errors'' appear as fabricational errors in
$H_8$.  Rather than having to deal with such errors during runtime, which is
computationally expensive and must be done in the circuit model, these
errors may instead be dealt with during the fabrication of $H_8$ in a much
more controlled environment.  Decoherence and other environmentally-induced
noise processes will still be present, but at least they will not be
conflated with dynamical control errors.

\begin{figure}
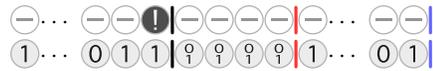

\begin{tabular}{lc@{}c@{}c@{}}
 & \nothing $\cdots$ \nothing \nothing \gatecursor \firstboundary & \nothing \nothing \nothing \nothing \secondboundary & \nothing $\cdots$ \nothing \nothing \thirdboundary\\
 & \one $\cdots$ \zero \one \one \firstboundary & \zeroone \zeroone \zeroone \zeroone \secondboundary & \one$\cdots$ \zero \one \thirdboundary
\end{tabular}
\caption{Start State: $\cdots$ represents an arbitrary number of qubit/qudit
vertical pairs}
\label{fig:start_state}
\end{figure}

%

\section{Universal quantum computation via adiabatic evolution to $H_8$}
\label{sec:adiabatic}

Given that comparable physical models have appeared in proofs for adiabatic
universality \cite{Aharonov:2004a, Aharonov:2007a}, we will
demonstrate that one-dimensional Hamiltonians on 8-level quantum systems on
a ring are universal under adiabatic evolution.  Our derivation will closely
mirror these previous works and follows straightforwardly from results of
Aharanov \etal\ \cite{Aharonov:2004a} that have been further refined by
Deift \etal\ \cite{Deift:2007a}.  We begin by quoting the adiabatic theorem:

\textbf{The Adiabatic Theorem (adapted from \cite{Deift:2007a}, quoted from
\cite{Aharonov:2004a})} Let $H_\textrm{init}$ and $H_\textrm{final}$ be two
Hamiltonians acting on a quantum system and consider the time-dependent
Hamiltonian $H(s) = (1-s)H_\textrm{init} + sH_\textrm{final}$.  Assume that
for all $s$, $H(s)$ has a unique ground state.  Then for any fixed $\delta >
0$, if
\begin{equation}
\label{eq:adiabatic}
T \geq
\Omega\left(\frac{||H_\textrm{final}-H_\textrm{init}||^{1+\delta}}{\epsilon^\delta\min_{s\in[0,1]}\left\{\Delta^{2+\delta}(H(s))\right\}}\right),
\end{equation}
then the final state of an adiabatic evolution according to $H$ for time $T$
(with an appropriate setting of global phase) is $\epsilon$-close in
$l_2$-norm to the ground state of $H_\textrm{final}$. The matrix norm is the
spectral norm $||H||=\max_w||Hw||/||w||$.

Loosely speaking, the adiabatic theorem states that if the variation between
the Hamiltonians is slow enough, then the quantum state will be in the ground
state of $H_\textrm{final}$ at the end of evolution if it starts in the
ground state of $H_\textrm{initial}$.  Consequently, to demonstrate that our
model is universal under adiabatic evolution, we must pick an
$H_\textrm{init}$ and $H_\textrm{final}$ whose ground states are
appropriately related to the initial and final states of computation.  

First consider $H_\textrm{init}$. We have previously detailed the
configuration of a valid initial state, $\ket{\psi_0}$, which encodes both the \emph{program
of execution} and the \emph{input data state}.  We define $H_\textrm{init}$
to be a sum over local projectors for which $\ket{\psi_0}$ is the
zero-eigenvalue ground state.  This differs from previous adiabatic proofs involving non-programmable architectures, which only needed to encode the input data state.  We will now
construct $H_\textrm{init}$ piece by piece. 

In a shorthand notation analagous to the one used in Sec.\
\ref{sec:quantum-walk-universality}, the first term is
\begin{equation}
I -
\ket{\begin{array}{c}
\nothing\gatecursor\\
\one\one
\end{array}}\!\!
\bra{\begin{array}{c}
\nothing\gatecursor\\
\one\one
\end{array}}_\firstboundary
\end{equation}
where the projector is for the two locations to the left of the swap
program/data region boundary.  While this will ensure that the ground state
is an initial state, it does not ensure that it is valid or that is executes
the desired program (i.e. that it is $\ket{\psi_0}$).  However, we can pick out
our desired initial state by adding nearest-neighbor projectors to this
Hamiltonian in a chained fashion.  The first such projector will be between
the leftmost qubit in the above projector and its left neighbor.  Without
loss of generality, suppose that the program qubit for our desired program
is a \zero\ in this location; we would then add terms
\begin{align}
&\sum_{\arbitraryqudit}
\ket{\begin{array}{c}
\arbitraryqudit\nothing\\
\one\one
\end{array}}\!\!
\bra{\begin{array}{c}
\arbitraryqudit\nothing\\
\one\one
\end{array}}_{\firstboundary-1}
\\
+
&\sum_{\arbitraryqudit \neq \nothing}
\ket{\begin{array}{c}
\arbitraryqudit\nothing\\
\zero\one
\end{array}}\!\!
\bra{\begin{array}{c}
\arbitraryqudit\nothing\\
\zero\one
\end{array}}_{\firstboundary-1}
\end{align}
where the first sum is over all qudit symbols, ensuring that anything over a
\one\ has a higher energy.  Similarly, the second sum ensures that all
configurations except the desired \nothing\ over a \zero\ have a higher
energy.  Thus, we sum over all non-desired configurations, of which there
are seven.  This process is continued by sliding over one location and again
summing over projectors where the right qudit-qubit pair is fixed to the
desired initial configuration and the left qudit-qubit pair runs over all
seven
illegal or non-desired configurations. Moving around the entire ring step by
step, we add similar projectors which tack on all undesired nearest-neighbor
pairs.  This ensures that the ground state is precisely $\ket{\psi_0}$; it is both valid and encodes only
the program we want to execute.  

Now consider $H_{\text{final}}$.  As discussed in \cite{Aharonov:2004a, Irani:2007a,
Aharonov:2007a,Deift:2007a}, we seek a $H_\text{final}$ whose ground state is the
sum-over-histories state
\begin{equation}
\frac{1}{\sqrt{\overline{T}+1}}\sum_{j=0}^{\overline{T}}|\psi_j\>.
\end{equation}
This can be accomplished simply by adding penalty terms to Eq.\
\ref{eq:HeffectiveWalk} which ``prefer'' the transition elements.  That is,
we define
\begin{align} 
H_{\text{final}} &:= \sum_{i=1}^{\overline{T}-1} |\psi_i\>\< \psi_i| \\
&-\frac{1}{2}\sum_{i=0}^{\overline{T}} |\psi_t\>\<\psi_{t-1}| + |\psi_{t-1}\>\<\psi_t|\\
& + \frac{1}{2}|\psi_0\>\<\psi_0| +
 \frac{1}{2}|\psi_{\overline{T}}\>\<\psi_{\overline{T}}|.
 \end{align}

Expressed in the $\ket{\psi_t}$ basis, $H_\textrm{init}$ is simply
\begin{equation}
H_\textrm{init} = 
\begin{pmatrix}
0 & 0 & \ldots & 0\\
0 & 1 & \ldots & 0\\
\vdots & \vdots & \ddots & \vdots\\
0 & 0 & \ldots & 1
\end{pmatrix}
\end{equation}
as $\ket{\psi_0}$ is the unique groundstate. Similarly, $H_\textrm{final}$
is
\begin{equation}
H_\textrm{final} = 
\begin{pmatrix}
 \frac{1}{2} & -\frac{1}{2} & 0            &   & \cdots &      & 0\\
-\frac{1}{2} & 1            & -\frac{1}{2} & 0      &\ddots&&\vdots\\
0            & -\frac{1}{2} & 1            & -\frac{1}{2} & 0&\ddots&\vdots\\
&\ddots&\ddots&\ddots&\ddots&\ddots&\\
\vdots&&0&-\frac{1}{2}&1&-\frac{1}{2}&0\\
&&&0&-\frac{1}{2}&1&-\frac{1}{2}\\
0&&\cdots&&0&-\frac{1}{2}&\frac{1}{2}
\end{pmatrix}.
\end{equation}
As proved in Sec 3.1.2 of \cite{Aharonov:2004a} and simplified in
\cite{Deift:2007a}, the spectral gap of these Hamiltonians is at least
$1/[2(\overline{T}+1)^2]$, which is an inverse polynomial in $T$, the number of
gates in the initial quantum circuit (recall $\overline{T}$ is polynomial in
$T$).  Thus, the evolution can be considered efficient, proving that
Hamiltonians of 8-level quantum systems on a one-dimensional ring are
universal for adiabatic quantum computation.

%

\section{A two-dimensional cylinder qubit Hamiltonian $H_2$}
\label{sec:qubits}

Given the conceptual decomposition of the 8-level quantum system into a
4-level quantum system and a qubit, it is easy to imagine fully decomposing
each 8-level quantum system into 3 qubits.  Such an architecture would still
be effectively one-dimensional, as the width of the ring is invariant.
However, each nearest neighbor interaction in the 8-level system would
become a 6-body interaction, which is a greater than Feynman's 4-body
interactions.

Alternatively, one could use gadget Hamiltonian theory, as presented in
\cite{Kempe:2004a}, to reduce the degree of interactions between qubits.
This procedure augments the original system $\cal{S}$ with additional
mediator qubits.  Given an initial Hamiltonian $H$ acting on $\cal{S}$, one
constructs a new Hamiltonian $H'$ which acts on the augmented system and
whose action restricted to $\cal{S}$ is bounded as $||H'-H||<\epsilon$ for a
desired $\epsilon$.  For our purposes, we would seek to replace the 6-body
terms required for the transition rules with lower body terms.  While in
theory it is straightforward to apply a 6-body gadget, each term in our
original Hamiltonian will require 6 mediators.  Generally, each 6-body term
will then require more than 6 mediators, resulting in an expanding geometry.
We have been unable to devise a scheme which maintains both spatially local
interactions and an effective one-dimensional width under this replacement
scheme.

An alternative and more promising approach, presented by Oliveira and Terhal
\cite{Oliveira:2005a}, details an ``efficient" gadget reduction procedure to
a 2-local Hamiltonian which is planar, but two dimensional.  We defer to the
paper for the details of such a reduction, but note that there is a
prescription for applying gadgets to reduce a spatially-sparse $k$-local
Hamiltonian to a 2-local Hamiltonian on a regular lattice.  This reduction
maintains the spatial setup of the initial Hamiltonian, so that for our ring
geometry, the reduced setup would be a regular lattice on the surface of a
cylinder.  Moreover, the effective interaction is not only close to the
initial Hamiltonian with respect to its eigenvalues, but the operator norm
is close within a restricted subspace.  Thus the dynamics are similarly
close.  In other words, given a desired $\epsilon$, a ``reduced'' 2-local
version of the architecture can be designed with dynamics which are
$\epsilon$-close to that of the original architecture.  While the actual
mapping requires fine-tuning perturbation coupling parameters to ensure only
polynomial overhead and growth, the Oliveira-Terhal approach does provide a
means for reducing the interaction degree of our architecture, though the
resulting system is on the surface of a cylinder whose height is no longer
fixed.

%

\section{Conclusion}
\label{sec:conclusion}

We presented a family of time-independent Hamiltonians that can enable
universal quantum computation either by driving a continuous-time quantum
walk or by terminating an adiabatic algorithm.  When used to drive a
continuous-time quantum walk, quantum computation consists of $1$) preparing
an input that describes a quantum circuit to be executed and the quantum
data onto which it is to be applied, $2$) waiting the appropriate amount of
time, and $3$) measuring the output.  The simplicity of the operation of
such a machine is appealing, but Feynman's original proposal for realizing
it \cite{Feynman:1986a} still remains far out of technological reach.  Our
work demonstrates that it suffices for such a machine to use only
nearest-neighbor interactions between 8-level systems on a 1D ring, which
may be more technologically feasible.  It also demonstrates that simulating
the dynamics of 1D time-independent Hamiltonians on 8-level systems is a
BQP-complete problem, even though it is known how to simulate 1D spin
systems in a way that scales polynomially with the number of spins
\cite{Osborne:2006a}.

When used to terminate an adiabatic algorithm, our Hamiltonian achieves
universality via nearest-neighbor interactions between 8-level systems on a
1D ring, one level fewer than in a recent proof of universality of the
adiabatic algorithm by 9-level nearest-neighbor 1D Hamiltonians
\cite{Aharonov:2007a}.

Finally, using gadget perturbation theory \cite{Oliveira:2005a}, our
Hamiltonian can be made spatially local using only qubits rather than
8-level systems, but at the expense of requiring the qubits to lie on the
two-dimensional geometry of a cylinder rather than a one-dimensional
geometry of a ring. 

A remaining challenge is to address error correction and fault tolerance
for adiabatic and quantum walk models driven by the Hamiltonian we
constructed \cite{Lidar:2007a}.  In particular, because the quantum walk
describing quantum computation is effectively a continuous-time quantum walk
on a line, an imperfect implementation may be subject to Anderson
localization which could exponentially suppress propagation along this line.
One promising feature of the quantum walk model is that there are no
dynamical controls during the operation of the quantum computer, so that
control errors are all fabricational.  This allows such errors to be handled
at ``compile time'' rather than at ``run time,'' which could be much easier.

Another remaining challenge is to explore whether finding the ground state
of our Hamiltonian is a QMA complete problem.  We conjecture that it is, as
several similar Hamiltonians used to prove adiabatic quantum computing
universality are \cite{Aharonov:2004a, Aharonov:2007a}.  However, the
``clock'' and its update rule as implemented by our Hamiltonian are not
localized to a particular point in space, as computation winds around the
ring many times during the course of a computation.  Hence, previous proofs
do not translate directly and creative ideas are required.  Since our
architecture is also programmable, the need to encode the program in the
initial state further complicates the QMA question.

\emph{Addendum}: As we were finishing this paper, we became aware of related
work by Nagaj and Wocjan that demonstrates that translationally invariant
quantum walks in one dimension (``continuous-time quantum cellular
automata'') are also universal for quantum computation, albeit using
ten-dimensional rather than eight-dimensional systems \cite{Nagaj:2008a}.  A
final remaining challenge we state is to explore whether our model can be
made translationally invariant or whether the dimension of the Nagaj-Wocjan
model can be reduced to eight as ours is.
 
%
\begin{acknowledgments}
AJL acknowledges support for this research from the Army Research Office
under contract W911NF-04-1-0242 and by the National Science Foundation under
contracts PHY-0555573 and PHY-0653596.
BAC acknowledges support for this research from a UNM-LANL Partnership in
Quantum Information Science Fellowship.
\end{acknowledgments}

%

\begin{widetext}

\begin{appendix}
\section{Example computation}
\label{appendix:example}

The following illustrates computing the circuit in Figure \ref{fig:circuit}. 
\begin{figure}[ht]
\[
\Qcircuit @C=.7em @R=.4em @! {
\lstick{\ket{\dataA}} & \qw &   \qw    &\gate{S}\\
\lstick{\ket{\dataB}} & \ctrl{1}&\qw   & \qw\\
\lstick{\ket{\dataC}} & \gate{S}   &\gate{H} & \ctrl{-2}
}
\]
\caption{Equivalent Circuit}
\label{fig:circuit}
\end{figure}
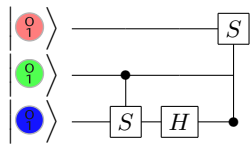

\begin{tabular}{llc@{}c@{}c@{}}
Start state&1.&\nothing \nothing \nothing \nothing \gatecursor \firstboundary & \nothing \nothing \nothing \secondboundary & \nothing \nothing \nothing \nothing \thirdboundary\\
&&\zero \zero \zero \one \one \firstboundary & \dataB \dataA \dataC \secondboundary & \one \zero \zero \zero \thirdboundary\\
\cline{3-5}
Swap \emph{data}&2.&\nothing \nothing \nothing \nothing \nothing \firstboundary & \gatecursor \nothing \nothing \secondboundary & \nothing \nothing \nothing \nothing \thirdboundary\\
&&\zero \zero \zero \one \one \firstboundary & \dataB \dataA \dataC \secondboundary & \one \zero \zero \zero \thirdboundary\\
\cline{3-5}
$\Lambda(S)$ on \dataB\dataC&3.&\nothing \nothing \nothing \nothing \nothing \firstboundary & \nothing \gatecursor \nothing \secondboundary & \nothing \nothing \nothing \nothing \thirdboundary\\
&&\zero \zero \zero \one \one \firstboundary & $\dataA \overline{\dataB} \overline{\dataC} $\secondboundary & \one \zero \zero \zero \thirdboundary\\
\cline{3-5}
$H$ since control bit is \one&4.&\nothing \nothing \nothing \nothing \nothing \firstboundary & \nothing \nothing \gatecursor\secondboundary & \nothing \nothing \nothing \nothing \thirdboundary\\
&&\zero \zero \zero \one \one \firstboundary & \dataA \dataB \dataC \secondboundary & \one \zero \zero \zero \thirdboundary\\
\cline{3-5}
Cycle \emph{Hadamard}&5.&\nothing \nothing \nothing \nothing \nothing \firstboundary & \nothing \nothing \nothing \secondboundary & \gatecursor \nothing \nothing \nothing \thirdboundary\\
&&\zero \zero \zero \one \one \firstboundary & \dataA \dataB $\overline{\dataC}$ \secondboundary & \one \zero \zero \zero \thirdboundary\\
\cline{3-5}
 &6.&\nothing \nothing \nothing \nothing \nothing \firstboundary & \nothing \nothing \nothing \secondboundary & \nothing \gatecursor \nothing \nothing \thirdboundary\\
&&\zero \zero \zero \one \one \firstboundary & \dataA \dataB \dataC \secondboundary & \zero \one \zero \zero \thirdboundary\\
\cline{3-5}
 &7.&\nothing \nothing \nothing \nothing \nothing \firstboundary & \nothing \nothing \nothing \secondboundary & \nothing \nothing \gatecursor\nothing \thirdboundary\\
&&\zero \zero \zero \one \one \firstboundary & \dataA \dataB \dataC \secondboundary & \zero \zero \one \zero \thirdboundary\\
\cline{3-5}
 &8.&\nothing \nothing \nothing \nothing \nothing \firstboundary & \nothing \nothing \nothing \secondboundary & \nothing \nothing \nothing \gatecursor \thirdboundary\\
&&\zero \zero \zero \one \one \firstboundary & \dataA \dataB \dataC \secondboundary & \zero \zero  \zero \one\thirdboundary\\
\cline{3-5}
Cycle \emph{swap}&9.&\gatecursor \nothing \nothing \nothing \nothing \firstboundary & \nothing \nothing \nothing \secondboundary & \nothing \nothing \nothing \nothing \thirdboundary\\
&&\zero \zero \zero \one \one \firstboundary & \dataA \dataB \dataC \secondboundary & \zero \zero \zero \one \thirdboundary\\
\cline{3-5}
&10.&\nothing \gatecursor \nothing \nothing \nothing \firstboundary & \nothing \nothing \nothing \secondboundary & \nothing \nothing \nothing \nothing \thirdboundary\\
&&\zero \zero \zero \one \one \firstboundary & \dataA \dataB \dataC \secondboundary & \zero \zero \zero \one \thirdboundary\\
\cline{3-5}
&11.&\nothing \nothing \gatecursor \nothing \nothing \firstboundary & \nothing \nothing \nothing \secondboundary & \nothing \nothing \nothing \nothing\thirdboundary\\
&&\zero \zero \zero \one \one \firstboundary & \dataA \dataB \dataC \secondboundary & \zero \zero \zero \one \thirdboundary\\
\cline{3-5}
&12.&\nothing \nothing \nothing \gatecursor \nothing \firstboundary & \nothing \nothing \nothing \secondboundary & \nothing \nothing \nothing \nothing \thirdboundary\\
&&\zero \zero \one \zero \one \firstboundary & \dataA \dataB \dataC \secondboundary & \zero \zero \zero \one \thirdboundary\\
\cline{3-5}
Become \cycle\ across boundary&13.&\nothing \nothing \nothing \nothing \gatecursor \firstboundary & \nothing \nothing \nothing \secondboundary & \nothing \nothing \nothing \nothing \thirdboundary\\
&&\zero \zero \one \one \zero \firstboundary & \dataA \dataB \dataC \secondboundary & \zero \zero \zero \one \thirdboundary\\
\cline{3-5}
Cycle \emph{data}&14.&\nothing \nothing \nothing \nothing \nothing \firstboundary & \cycle \nothing \nothing \secondboundary & \nothing \nothing \nothing \nothing \thirdboundary\\
&&\zero \zero \one \one \zero \firstboundary & \dataA \dataB \dataC \secondboundary & \zero \zero \zero \one \thirdboundary\\
\cline{3-5}
&15.&\nothing \nothing \nothing \nothing \nothing \firstboundary & \nothing \cycle \nothing \secondboundary & \nothing \nothing \nothing \nothing \thirdboundary\\
&&\zero \zero \one \one \zero \firstboundary & \dataB \dataA \dataC \secondboundary & \zero \zero \zero \one \thirdboundary\\
\cline{3-5}
&16.&\nothing \nothing \nothing \nothing \nothing \firstboundary & \nothing \nothing \cycle \secondboundary & \nothing \nothing \nothing \nothing  \thirdboundary\\
&&\zero \zero \one \one \zero \firstboundary & \dataB \dataC \dataA & \secondboundary \zero \zero \zero \one \thirdboundary\\
\cline{3-5}
Cycle \emph{Hadamard}&17.&\nothing \nothing \nothing \nothing \nothing \firstboundary & \nothing \nothing \nothing \secondboundary & \cycle \nothing \nothing \nothing \thirdboundary\\
&&\zero \zero \one \one \zero \firstboundary & \dataB \dataC \dataA \secondboundary & \zero \zero \zero \one \thirdboundary\\
\cline{3-5}
&18.&\nothing \nothing \nothing \nothing \nothing \firstboundary & \nothing \nothing \nothing \secondboundary & \nothing \cycle \nothing \nothing \thirdboundary\\
&&\zero \zero \one \one \zero \firstboundary & \dataB \dataC \dataA \secondboundary & \zero \zero \zero \one \thirdboundary\\
\cline{3-5}
&19.&\nothing \nothing \nothing \nothing \nothing \firstboundary & \nothing \nothing \nothing \secondboundary & \nothing \nothing  \cycle \nothing \thirdboundary\\
&&\zero \zero \one \one \zero \firstboundary & \dataB \dataC \dataA \secondboundary & \zero \zero \zero \one \thirdboundary\\
\cline{3-5}
&20.&\nothing \nothing \nothing \nothing \nothing \firstboundary & \nothing \nothing \nothing \secondboundary & \nothing \nothing \nothing \cycle\thirdboundary\\
&&\zero \zero \one \one \zero \firstboundary & \dataB \dataC \dataA \secondboundary & \zero \zero \one \zero\thirdboundary\\
\cline{3-5}
Cycle \emph{swap}&21.&\cycle \nothing \nothing \nothing \nothing \firstboundary & \nothing \nothing \nothing \secondboundary & \nothing \nothing \nothing \nothing \thirdboundary\\
&&\zero \zero \one \one \zero \firstboundary & \dataB \dataC \dataA \secondboundary & \zero \zero \one \zero \thirdboundary\\
\cline{3-5}
&22.&\nothing \cycle \nothing \nothing \nothing \firstboundary & \nothing \nothing \nothing \secondboundary & \nothing \nothing \nothing \nothing\thirdboundary\\
&&\zero \zero \one \one \zero \firstboundary & \dataB \dataC \dataA \secondboundary & \zero \zero \one \zero \thirdboundary\\
\cline{3-5}
&23.&\nothing \nothing \cycle \nothing \nothing \firstboundary & \nothing \nothing \nothing \secondboundary & \nothing \nothing \nothing \nothing \thirdboundary\\
&&\zero \one \zero \one \zero \firstboundary & \dataB \dataC \dataA \secondboundary & \zero \zero \one \zero \thirdboundary\\
\cline{3-5}
&24.&\nothing \nothing \nothing \cycle \nothing \firstboundary & \nothing \nothing \nothing \secondboundary & \nothing \nothing \nothing \nothing \thirdboundary\\
&&\zero \one \one \zero \zero \firstboundary & \dataB \dataC \dataA \secondboundary & \zero \zero \one \zero \thirdboundary\\
\cline{3-5}
Become \holdcycle\ across boundary&25.&\nothing \nothing \nothing \nothing \cycle \firstboundary & \nothing \nothing \nothing \secondboundary & \nothing \nothing \nothing \nothing \thirdboundary\\
&&\zero \one \one \zero \zero \firstboundary & \dataB \dataC \dataA \secondboundary & \zero \zero \one \zero \thirdboundary\\
\cline{3-5}
\end{tabular}

\begin{tabular}{llc@{}c@{}c@{}}
Swap \emph{data}&26.&\nothing \nothing \nothing \nothing \nothing \firstboundary & \holdcycle \nothing \nothing \secondboundary & \nothing \nothing \nothing \nothing \thirdboundary\\
&&\zero \one \one \zero \zero \firstboundary & \dataB \dataC \dataA \secondboundary & \zero \zero \one \zero\thirdboundary\\
\cline{3-5}
Do not swap \emph{data}&27.&\nothing \nothing \nothing \nothing \nothing \firstboundary & \nothing \holdcycle \nothing \secondboundary & \nothing \nothing \nothing \nothing \thirdboundary\\
&&\zero \one \one \zero \zero \firstboundary & \dataC \dataB \dataA \secondboundary & \zero \zero \one \zero \thirdboundary\\
\cline{3-5}
&28.&\nothing \nothing \nothing \nothing \nothing \firstboundary & \nothing \nothing \holdcycle \secondboundary & \nothing \nothing \nothing \nothing \thirdboundary\\
&&\zero \one \one \zero \zero \firstboundary & \dataC \dataB \dataA \secondboundary & \zero \zero \one \zero \thirdboundary\\
\cline{3-5}
Cycle \emph{Hadamard}&29.&\nothing \nothing \nothing \nothing \nothing \firstboundary & \nothing \nothing \nothing \secondboundary & \holdcycle \nothing \nothing \nothing \thirdboundary\\
&&\zero \one \one \zero \zero \firstboundary & \dataC \dataB \dataA \secondboundary & \zero \zero \one \zero\thirdboundary\\
\cline{3-5}
&30.&\nothing \nothing \nothing \nothing \nothing \firstboundary & \nothing \nothing \nothing \secondboundary & \nothing \holdcycle \nothing \nothing \thirdboundary\\
&&\zero \one \one \zero \zero \firstboundary & \dataC \dataB \dataA \secondboundary & \zero \zero \one \zero\thirdboundary\\
\cline{3-5}
&31.&\nothing \nothing \nothing \nothing \nothing \firstboundary & \nothing \nothing \nothing \secondboundary & \nothing \nothing \holdcycle \nothing \thirdboundary\\
&&\zero \one \one \zero \zero \firstboundary & \dataC \dataB \dataA \secondboundary & \zero \one   \zero \zero\thirdboundary\\
\cline{3-5}
&32.&\nothing \nothing \nothing \nothing \nothing \firstboundary & \nothing \nothing \nothing \secondboundary & \nothing  \nothing \nothing \holdcycle \thirdboundary\\
&&\zero \one \one \zero \zero \firstboundary & \dataC \dataB \dataA \secondboundary & \zero \one \zero \zero\thirdboundary\\
\cline{3-5}
Cycle \emph{swap} &33.&\holdcycle \nothing \nothing \nothing \nothing \firstboundary & \nothing \nothing \nothing \secondboundary & \nothing \nothing \nothing \nothing \thirdboundary\\
&&\zero \one \one \zero \zero \firstboundary & \dataC \dataB \dataA \secondboundary & \zero \one \zero \zero \thirdboundary\\
\cline{3-5}
&34.&\nothing \holdcycle \nothing \nothing \nothing \firstboundary & \nothing \nothing \nothing \secondboundary & \nothing \nothing \nothing \nothing \thirdboundary\\
&&\one \zero \one \zero \zero \firstboundary & \dataC \dataB \dataA \secondboundary & \zero \one \zero \zero \thirdboundary\\
\cline{3-5}
&35.&\nothing \nothing \holdcycle \nothing \nothing \firstboundary & \nothing \nothing \nothing \secondboundary & \nothing \nothing \nothing \nothing \thirdboundary\\
&&\one \one \zero \zero \zero \firstboundary & \dataC \dataB \dataA \secondboundary & \zero \one \zero \zero\thirdboundary\\
\cline{3-5}
&36.&\nothing \nothing \nothing \holdcycle \nothing \firstboundary & \nothing \nothing \nothing \secondboundary & \nothing \nothing \nothing \nothing\thirdboundary\\
&&\one \one \zero \zero \zero \firstboundary & \dataC \dataB \dataA \secondboundary & \zero \one \zero \zero\thirdboundary\\
\cline{3-5}
Become \gatecursor\ across boundary&37.&\nothing \nothing \nothing \nothing \holdcycle \firstboundary & \nothing \nothing \nothing \secondboundary & \nothing \nothing \nothing \nothing\thirdboundary\\
&&\one \one \zero \zero \zero \firstboundary & \dataC \dataB \dataA \secondboundary & \zero \one \zero \zero\thirdboundary\\
\cline{3-5}
Swap \emph{data}&38.&\nothing \nothing \nothing \nothing \nothing \firstboundary & \gatecursor \nothing \nothing \secondboundary & \nothing \nothing \nothing \nothing \thirdboundary\\
&&\one \one \zero \zero \zero \firstboundary & \dataC \dataB \dataA \secondboundary & \zero \one \zero \zero\thirdboundary\\
\cline{3-5}
$\Lambda(S)$ on \dataC\dataA &39.&\nothing \nothing \nothing \nothing \nothing \firstboundary & \nothing \gatecursor \nothing \secondboundary & \nothing \nothing \nothing \nothing\thirdboundary\\
&&\one \one \zero \zero \zero \firstboundary &  \dataB \dataC \dataA \secondboundary & \zero \one \zero \zero \thirdboundary\\
\cline{3-5}
No $H$ since control bit is \zero&40.&\nothing \nothing \nothing \nothing \nothing \firstboundary & \nothing \nothing \gatecursor \secondboundary & \nothing \nothing \nothing \nothing\thirdboundary\\
&&\one \one \zero \zero \zero \firstboundary & \dataB $\overline{\dataC} \overline{\dataA}$ \secondboundary & \zero \one \zero \zero\thirdboundary\\
\cline{3-5}
Cycle Control Program&41.&\nothing \nothing \nothing \nothing \nothing \firstboundary & \nothing \nothing \nothing \secondboundary & \gatecursor \nothing \nothing \nothing  \thirdboundary\\
&&\one \one \zero \zero \zero \firstboundary & \dataB \dataC \dataA
\secondboundary & \zero \one \zero \zero \thirdboundary \\
%
\mbox{Become \gatecursor\ across boundary}&42.&\nothing \nothing \nothing \nothing \nothing \firstboundary & \nothing \nothing \nothing \secondboundary & \nothing  \gatecursor \nothing \nothing  \thirdboundary\\
&&\one \one \zero \zero \zero \firstboundary & \dataB \dataC \dataA \secondboundary & \one \zero \zero \zero \thirdboundary\\
\cline{3-5}
&43.&\nothing \nothing \nothing \nothing \nothing \firstboundary & \nothing \nothing \nothing \secondboundary & \nothing \nothing \gatecursor \nothing  \thirdboundary\\
&&\one \one \zero \zero \zero \firstboundary & \dataB \dataC \dataA \secondboundary & \one \zero \zero \zero \thirdboundary\\
\cline{3-5}
&44.&\nothing \nothing \nothing \nothing \nothing \firstboundary & \nothing \nothing \nothing \secondboundary & \nothing \nothing \nothing  \gatecursor \thirdboundary\\
&&\one \one \zero \zero \zero \firstboundary & \dataB \dataC \dataA \secondboundary & \one \zero \zero \zero \thirdboundary\\
\cline{3-5}
Cycle \emph{swap}&45.&\gatecursor \nothing \nothing \nothing \nothing \firstboundary & \nothing \nothing \nothing \secondboundary & \nothing \nothing \nothing \nothing \thirdboundary\\
&&\one \one \zero \zero \zero \firstboundary & \dataB \dataC \dataA \secondboundary & \one \zero \zero \zero \thirdboundary\\
&46.&\nothing \gatecursor \nothing \nothing \nothing \firstboundary & \nothing \nothing \nothing \secondboundary & \nothing \nothing \nothing \nothing \thirdboundary\\
&&\one \one \zero \zero \zero \firstboundary & \dataB \dataC \dataA \secondboundary & \one \zero \zero \zero \thirdboundary\\
&47.&\nothing \nothing \gatecursor \nothing \nothing \firstboundary & \nothing \nothing \nothing \secondboundary & \nothing \nothing \nothing \nothing \thirdboundary\\
&&\one \zero \one \zero \zero \firstboundary & \dataB \dataC \dataA \secondboundary & \one \zero \zero \zero \thirdboundary\\
Stop state&48.& \nothing \nothing \nothing  \gatecursor\nothing \firstboundary & \nothing \nothing \nothing \secondboundary & \nothing \nothing \nothing \nothing \thirdboundary\\
&&\one \zero \zero \one  \zero \firstboundary & \dataB \dataC \dataA \secondboundary & \one \zero \zero \zero \thirdboundary\\
\end{tabular}
\end{appendix}
\end{widetext}

\bibliographystyle{\BstFile}
\bibliography{\BibFile}

\end{document}